# Deep learning based sferics recognition for AMT data processing in the dead band

Enhua Jiang[1,2,3], Rujun Chen[1,2,3,*], Xinming Wu[4,*], Jianxin Liu[1,2,3], Debin Zhu[1,2,3], Regean Pitiya[1,2,3], Weiqiang Liu[5]

[1] *Key Laboratory of Metallogenic Prediction of Nonferrous Metals and Geological Environment Monitoring, Ministry of Education（Central South University）, Changsha, China, 410083. E-mail: 205011075@csu.edu.cn*
[2] *Key Laboratory of Non-ferrous Resources and Geological Hazard Detection, Central South University, Changsha, China, 410083.*
[3] *School of Geoscience and Info-physics, Central South University, South Lushan Road, Changsha, China 410083.*
[4] *School of Earth and Space Sciences, University of Science and Technology of China, Hefei 230026, China*
[5] *Chinese Academy of Geological Sciences, Beijing, 100037, China*
*Correspondence: chrujun@csu.edu.cn (R.C.); xinmwu@ustc.edu.cn (X.W.)

## SUMMARY

The main signals from the field sources of audio-magnetotellurics (AMT) sounding generated by global lightning activities are called sferics. In data analysis, the absence of long-term and high-amplitude natural field signals will result in a lack of energy in the AMT dead-band, which may cause unreliable impedance estimate. The collected data itself is redundant, and most of the data segments are non-signal segments containing noise. To achieve reliable AMT data interpretation, professional and rigorous preprocessing methods such as cropping or selection are required. In order to efficiently extract AMT main signals, we propose a deep convolutional neural network (CNN) to automatically recognize sferic signals from time series. We train the CNN by using field time series data with different signal to noise rations (S/Ns) that were acquired from different regions in mainland China. To solve the potential overfitting problem due to the limited number of sferic labels, we propose a training strategy that randomly generates training samples (with random data augmentations) while optimizing the CNN model parameters. We stop the process of training and data generation until the training loss converges. In addition, we use a weighted binary cross-entropy loss function to solve the sample imbalance problem to better optimize the network, use multiple reasonable metrics to evaluate network performance, and carry out ablation experiments to optimally choose the model hyperparameters. Extensive field data applications show that our trained CNN can robustly recognize sferic signals from noisy time series for subsequent impedance estimation. The computed results show that our method can significantly improve S/N and effectively solve the problem of lack of energy in AMT dead-band. We obtained a smoother and more reasonable apparent resistivity-phase curves and depolarized phase tensor in the dead-band. corrected the estimation error of sudden drop/abrupt change of high-frequency apparent resistivity and abnormal behavior of phase reversal (presented by pseudo-sections), and finally restored the real shallow subsurface resistivity structure.

## I. INTRODUCTION

In AMT data processing, using data with a low S/N in the horizontal field channel will lead to biased transfer function estimates (Labson et al., 1985). Therefore, field source investigation is of great significance for AMT sounding (Egbert, 1986; Garcia & Jones, 2002), particularly at dead-band frequencies (1.5–5 kHz). Several studies have shown that strong seasonal and diurnal variations in global lightning activity (Price, 1993; Chrissan & Fraser-Smith, 1996; Satori & Zieger, 1996; Fiillekrug & Fraser-Smith, 1997; Watkins et al., 1998). The magnetic field signal level is usually lower than the coil noise threshold during the day. In contrast, the signal level at night is usually strong enough to robustly estimate the transfer function at AMT dead-band frequencies (Garcia & Jones, 2002). Based on this observation, Garcia & Jones (2005) proposed a new hybrid method of acquisition and processing, which solves the problem of the lack of energy in the AMT dead-band at high latitudes.

Over the past few decades, much research has been devoted to sferic signals. The arrival time of sferics is related to the rate of thunderstorm dimensions (Cherna et al, 1986), then people can use the arrival time difference (ATD) technique to record the lightning activity (Lee, 1986). Such approaches, however, have failed to remove propagation effects from ground conductivity or ionospheric conductivity profiles. In order to alleviate the impact of this limitation, scholars have since proposed many new methods for precise positioning (Lee,1989). And all these ATD techniques form the basis of the World Wide Lightning Location Network (WWLLN) (Dowden et al., 2002; Christian, et al., 2003; Jacobson et al., 2006). Moreover, new WWLLNs and lightning positioning technologies are constantly being constructed and proposed (Anagnostou, 2006; Said et al., 2010).

The rapid development of WWLLN has promoted many studies using sferics to improve the S/N of AMT (Slankis et al.,1972; Tzanis & Beamish, 1987; Toledo-Redondo et al., 2010; Grandt,1991; Chave and Jones, 2012). By extracting the high-amplitude and close-range sferics provided in WWLLN, S/N can be significantly increased in the dead-band (Hennessy, L. et al., 2015). The amplitude, arrival time and azimuth of each sferic in the time series can be predicted by using lightning network data and modeling the propagation of EIWG. Then, A new method for processing and interpreting AMT data from natural fields can be developed by associating sferic with lightning source parameters (Hennessy, L. et al., 2018, 2019).

For noisy AMT data, the decomposition algorithms make use of statistical techniques such as the jackknifes and the bootstraps to constrain distorted models (Chave & Smith 1994; McNeice & Jones 2001), but do not take advantage of the high S/Ns data. With the development of AMT field source theory, much research has tended to focus on extracting limited sections of high S/Ns data for subsequent data processing (Garcia & Jones 2002) rather than on processing the poor-quality parts of data (i.e., improve S/N by precisely controlling extracted sferics with amplitudes above noise level observed in AMT time series). By averaging the extracted data at the same source time, S/N can be made proportional to the square root of the mean (Macnae et al., 1984; McCracken et al., 1986). Based on this finding, Goldak & Goldak (2001) intended to use adaptive polarization stacking to improve S/N, but failed to take into account the disturbance effect of average non-stationary sferic waveforms. Hennessy & Macnae (2018) found that data between sferics events is irrelevant and can be discarded, and if sferics are duplicated (or similar), theoretically, stacking them will improve S/N, as well as reducing the deviation in apparent resistivity and phase curves.

Initially, scholars used statistical information such as the sample mean and variance to calculate the 90% confidence interval of the average value to determine the number of sferic needed to reach a robust mean (Leon-Garcia, 1994). However, Short distance lightning activities can produce individual transient events, the amplitude of which are obviously larger than that of low-level background field, so the S/N can be significantly improved by recording a transient source. In addition to this, several studies have revealed that processing a single sferic event based on polarization, distance and amplitude can also increase the S/N, especially in the AMT dead-band (Garner & Thiel, 2000; Goldak & Goldak, 2001; Garcia & Jones, 2008; Hennessy & Macnae, 2018; Cong Zhou et al., 2021).

To date, researchers have investigated a variety of approaches to sferics recognition and extraction, but there is still no efficient solution. Currently the main method of precisely controlling the extraction of sferics relies on WWLLN, which can detect the arrival time, location and pulse of up to 4 million lightning strikes per day, as well as calculate and catalog the distribution of different regions (Grandt,1991; Ushio et al., 2015). Nevertheless, this method picks up sferics manually through station records, that is, by recording the time of lightning activity, then measuring its distance, and finally estimating the arrival time on the time series for recognition. This method often makes data processing process more complicated and greatly increases the labor and time costs.

In recent years, with the development of hardware, distributed systems and cloud computing technology, especially the progress of neural-network-based computing, Deep learning provides novel solutions to many recognition tasks including time series classification (Wang et al., 2017; Ismail et al., 2019; Paul A, 2020) and waveform picking (W Zhu & Beroza, 2019; Ardaillon & Roebel,2020; Yu & Ma, 2021). These methods use deep learning to achieve superior performance. And as long as the network is well-trained, it only takes a short time in the forward prediction. Inspired by these technologies, compared with the traditional method, this paper presents a new method for recognizing sferics, which discards the complex network positioning steps and is no longer limited by station equipment, making the AMT field exploration more convenient. Specifically, we use deep CNN to recognize sferics, extract the signal with the closest waveform and the highest amplitude, and discard the data between sferic events to increase the S/N. In order to verify the performance of our method, we have carried out extensive tests on different S/Ns measured data. And the experimental results indicated that the well-trained CNN model has achieved outstanding generalization and high computational efficiency. From the sferics recognized by our CNN model, we have obtained smooth and reasonable apparent resistivity and phase curves in the AMT dead-band, thus reflecting the real subsurface resistivity structure.

The remainder of this paper is structured as follows: first we start by introducing the network architecture, dataset production process and training parameter settings in Sect. 2. In Sect. 3, we describe/explain the principle and implementation of a robust impedance estimation algorithm for subsequent impedance calculations on/(data processing of) recognized sferic signals. In Sect. 4, we apply two field data applications to evaluate the proposed approach/method, and use the apparent resistivity and phase tensor to show/demonstrate the effectiveness and goof generalization of our approach/method. In Sect. 5, we discuss some possible problems and limitations. Finally, we present a conclusion in Sect. 6 to summarize our findings and discuss future directions.

## II. SFERIC SIGNALS RECOGNITION

In this section, we aim to accurately recognize sferic signals from AMT time series. To this end, we conduct an ablation experiment to decide which architecture of deep neural network is suitable for this task, and finally choose one-dimensional VGG19 as the training network. To properly supervise the network learning process, weighted BCE loss function is used to optimize the network parameters. In addition, various quality metrics are proposed to evaluate the predicted results and demonstrate the superior performance of the network.

Furthermore, we propose a data generation method that randomly generates training samples with random data augmentation at training/at the same time as training. By following this method, we effectively mitigate the potential overfitting problem caused by sample imbalance. Finally, we stop the process of training and data generation until the training loss converges.

**A. Network Design/Details**

The network in this paper is a one-dimensional variant of VGG. To obtain more realistic results, we use a weighted BCE loss function to reduce the effects of sample imbalances. Moreover, we propose multiple types of quality metrics to evaluate the predictions.

*Network architecture*

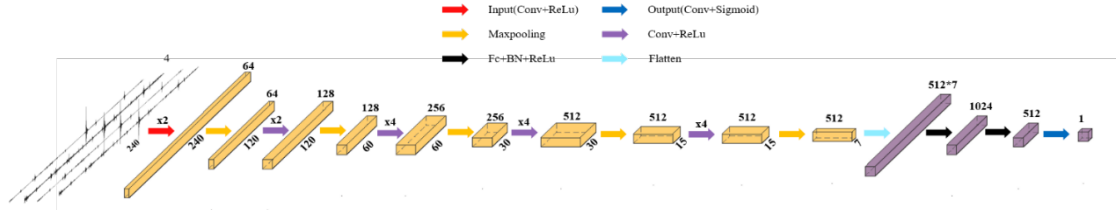

Fig. 1. Network architecture used in our proposed method.

The sferics recognition is a supervised learning task that typically requires a large amount of labeled data to obtain a network with excellent performance. However, it is almost impossible to make a complete point-by-point marker for an entire AMT time series, so the sferics recognition can be considered as a binary classification task. The architecture of our network is based on one-dimensional VGG (Simonyan et al**., 2014**). Due to its concise structure and deep network structure, VGG can control the number of parameters while obtaining more signal features, and has been widely used in many classification tasks. In this network, the inputs are mapped to the output of the classification category, where the features are expressed in low rank at the deepening network structure layer and then are fused through several fully connected layers. Lastly, the classification is completed through a fully connected layer.

The network architecture used in this paper is illustrated in Fig. 1, which consists of two parts: a feature extraction block and a classification block. The feature extraction block consists of five down-sampling blocks, each downs-ampling block consists of four convolutional layers with kernel 1×3, a max-pooling layer with kernel 1 × 2 and stride 2, and each convolution layer is followed by a rectified linear unit (ReLU) (Nair & Hinton, 2010). The feature extraction block extracts abstract data representations progressively by stacking such convolutional layers and pooling layers continuously, so as to enlarge the receptive fields of the convolutional layers and improve the performance of network feature extraction. The classification block consists of two feature aggregation blocks and a neuron for the final output. Each feature aggregation block starts with a fully connected layer, followed by a batch normalization layer (BN) (Ioffe & Szegedy, 2015) and ReLu.

Our network enriches the learned feature maps by continuously deepening the channels. The network uses fully connected layers to systematically aggregate multi-channel information. The fully connected layer aggregates losslessly the feature information extracted by convolutional layers, which helps to make use of global information for accurate prediction. The max pooling layer is better at capturing changes in the time step, bringing greater local information difference while maintaining translation invariance, in which a small pooling kernel can capture more detailed information and better describe sferics waveform features, etc.

*Loss functions*

Sferics recognition is a binary classification problem, in which the class of each sample is classified according to the corresponding fixed-length sampling points. In many binary classification problems, the binary cross-entropy loss (BCE) is the most common loss function used to measure the difference between the hypothesized class and the truth class.

First, in order to reflect probability distribution, network output needs to be mapped to the value range between 0 and 1 through the sigmoid activation function. Subsequent, the BCE loss is computed by the value after the sigmoid activation function. Where the Sigmoid activation function is defined as

$$O(x) = \frac{1}{1+e^{-x}} \quad (1)$$

Then, BCE loss can be mathematically defined as

$$l(x,y) = \{\mathcal{L}_1, \dots, \mathcal{L}_N\}^T \quad (2)$$

where $N$ is the batch size, $\mathcal{L}$ is defined as

$$\mathcal{L}_{BCE} = -\sum_{i=1}^{n} y_i \cdot log(O(x_i)) - \sum_{i=1}^{n}(1-y_i) \cdot log(1-O(x_i)) \quad (3)$$

where $x_i$ represents the output of the network and is converted to class probability after the sigmoid activation function $O(\cdot)$. $y_i$ represents the ground truth of binary labels, with a value of 0 or 1. $n$ represents the number of all samples. However, sferics recognition is a sample imbalance task, and this problem will force the network to be biased towards learning more non-sferic (negative sample) features, which will eventually lead to more true positive samples being predicted as negative samples. For subsequent impedance estimation, it will not be able to extract enough sferics to obtain robust results in the dead-band.

In order to solve the sample imbalance problem, we update the loss function to the weighted BCE loss. Mathematically, BCEW is defined as

$$\mathcal{L}_{BCEW} = -\beta \sum_{i=1}^{n} y_i \cdot log(O(x_i)) - (1-\beta) \sum_{i=1}^{n} (1-y_i) \cdot log(1-O(x_i)) \quad (4)$$

where $\beta$ represents the ratio between non-sferic samples and the total samples. $1-\beta$ represents the ratio of sferic samples.

*Quality metrics*

When training the network for sferics recognition, we use a series of metrics, including Accuracy, Precision, Recall and F1 score, to quantitatively evaluate the predicted results from multiple perspectives. The mathematical meaning of each symbol is shown in Table I

$$A = \frac{TP+TN}{TP+TN+FP+FN} \quad (5)$$

$$P = \frac{TP}{TP+FP} \quad (6)$$

$$R = \frac{TP}{TP+FN} \quad (7)$$

$$F1 = \frac{2PR}{P+R} \quad (8)$$

Table I
Confusion matrix for binary classification

| Truth class | Hypothesized class | |
|---|---|---|
| | True | False |
| Positive | TP | FN |
| Negative | FP | TN |

*$TPR = \frac{TP}{TP+FN}$, $FPR = \frac{FP}{TN+FP}$

where $A$ represents the ratio between correctly classified positive samples and the total samples, $P$ represents the ratio of the true positive samples in the predicted positive samples, $R$ represents the ratio of the predicted positive samples in the true positive samples, $F1$ represents a "balance point", which is a weighted harmonic average of P and R, with more focus on the lower of the two.

**B. Training Data Sets**

Before training a model for sferics recognition, we need tremendous amounts of AMT time series data for labeling. Considering the influence of different regions on the distribution of field source, we use the data from three surveys (i.e., Tibet (high S/N), Nanjing (medium S/N) and Wuhan (low S/N)) to make a training set. we follow a principle provided by Hennessy et al (2018) to select the window with amplitude significantly higher than the background noise level and wave shape closest to the large amplitude sferic as a positive sample. Besides, we implement a data augmentation scheme to further enlarge the training data set and improve the model generalization.

*Data generation*

To properly train the network, we develop a workflow (Fig. 2) to automatically build the sufficiently large training dataset with various waveform characteristics. For example, we denote a period of time series containing sferic as $sx$; and $ps$ is the position index of the sferic center point in the time series, which is used to make the sample mask $mx$; $mx$, initialized to an all-zero sequence of the same length as $sx$, takes $ps$ as the center point and assigns the sampling point with the extension radius $r$ on both sides to 1, which is used for sampling mask.

Different from the general data labeling, we use random sampling to generate positive and negative samples for training and validation sets in this workflow. Specifically, we set the sample length to $n$ and use a random sliding window to generate positive samples, as shown by the dashed window in Fig.3. Therefrom, we can generate at most $n - 2 * r$ positive samples from a single sferic in original time series, these positive samples contain all possible positions of sferics from left to right in the sample, as shown in the generated samples in Fig.3. On the one hand, by expanding the number of positive samples in this way, we aim to solve the sample imbalance problem to some extent, and on the other hand, it is beneficial for our model to learn the features of sferics at different positions within the sampling window.

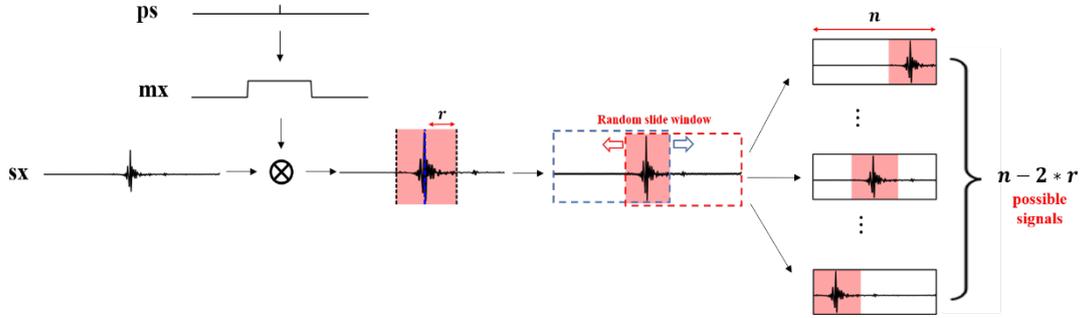

Fig. 2. To properly train the network and solve the sample imbalance problem, we develop a workflow to randomly generate multiple positive samples from a single sferic.

By following this workflow, we obtained significantly more positive samples than are contained in the original time series. To make the training data generally representative, here we use the time series data acquired from three different S/Ns surveys in Tibet, Nanjing, and Wuhan to make the training set. S/N is divided based on the number and amplitude of sferics contained within a unit time.

In our experiment, we divided 90 of the total 150 stations to contribute training set containing 30,000 positive samples, 30 stations for validation set containing 10,000 positive samples, and 30 stations for test set. According to the principle of stratified sampling, the samples ratio of the three surveys in each dataset is 6:2:2. The section below describes sferics waveform features on time series with different S/Ns.

Fig. 3(a) shows a fragment of a 100 ms time series from a station in Shannan, Tibet. From the graph we can see that local time series usually contains about 3-4 large-amplitude sferics and several clusters of small-amplitude sferics per 100 ms, which occur at random times with a duration of up to 1.5 ms. Fig. 3(b) extracts part of the time series in Fig. 3(a) to provide a detailed information, this time series segment contains very obvious large-amplitude sferics which have a well-structured waveform with amplitudes up to tens of millivolts, and well beyond the natural background sources level, The generation of such high S/N data is likely related to the local high altitude and distance from urban sources of interference, and these high-quality sferic signals are ideal positive samples in the training set. Fig. 3(c) shows a time series segment of a station in Nanjing, Jiangsu. This survey is located on the edge of the city and is disturbed by a certain degree of background cultural noise. By observing time series of the survey, we found that the acquired AMT time series is interfered by persistent noise sources (i.e., transformers and power lines). As same as Fig. 3(b), Fig. 3(d) shows more details in Fig. 3(c), these sferics are smaller in number and lower in amplitude than Tibet, which also reflects the diversity and complexity of sferics in various geological conditions. Fig. 3(e) shows a time series segment of a station in Wuhan, Hubei. Wuhan has a complex geological environment, and its interference sources are various and irregular. The acquired sferics have been almost submerged in complex EM noise, and also accompanied by strong interference such as periodic pulses during the acquisition period. It is hardly possible to distinguish whether there are sferics in this segment from Fig.3(f). However, such time series is one of the most common situations in AMT field sounding.

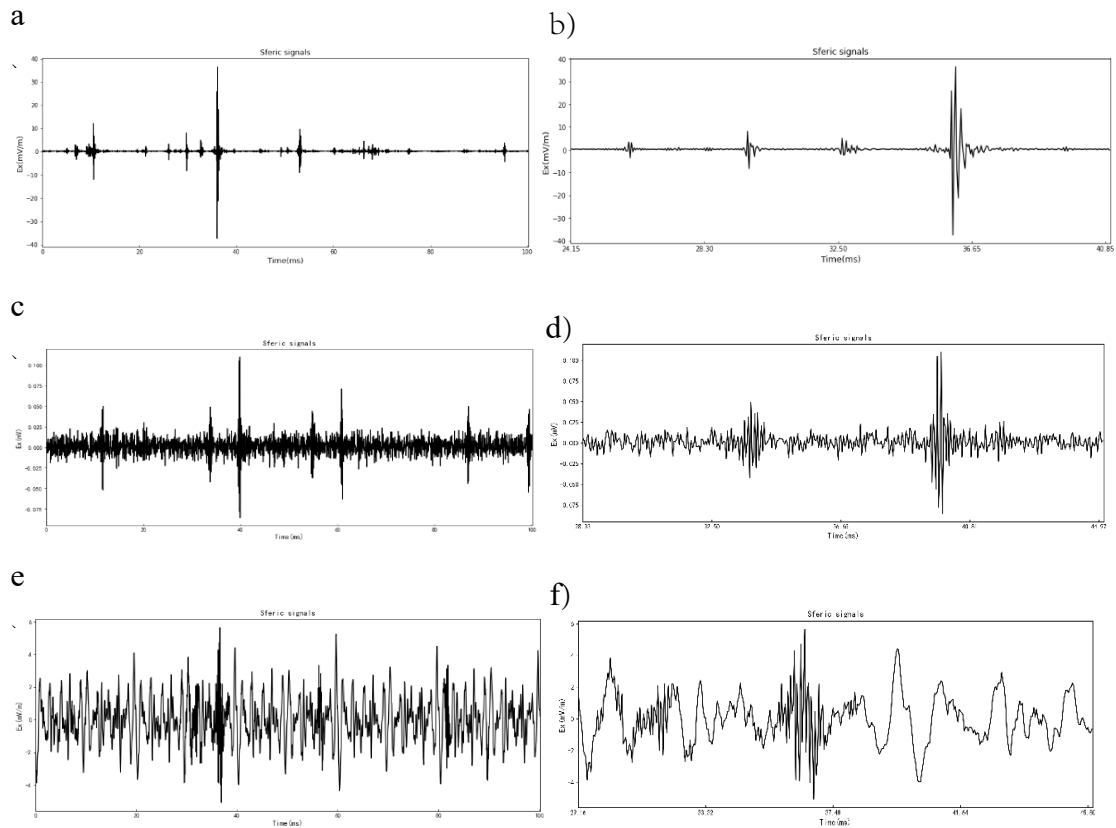

Fig. 3. We show a subset of training data with different S/Ns generated by the proposed workflow: (a) a segment of $E_x$ component data with high-amplitude sferics (black); (b) time period 20.00ms-39.98ms in (a); (c) a segment of $E_x$ component data with medium-amplitude sferics (black); (d) time period 31.16ms-43.38ms in (c); (e) a segment of $E_x$ component data with low-amplitude sferics (black); (f) time period 25.00ms-38.81ms in (e).

*Data augmentation*

Data augmentation is one of the most useful ways for improving the performance of deep models. There are some successful instances in geophysics (F. Li et al., 2021). After generating the dataset consisting of 90 stations, we employ a simple data augmentation to further increase the diversity of our training data set.

In order to satisfy the assumption of identically distributed data and to keep the sferics waveform uncorrupted, in generating the input time series data for training, we have added random noise into the data as shown in Fig. 4, where the added noise makes time series look more realistic. To increase the generalization of the training data while avoiding sferics being overwhelmed by added noise, the S/N for each training sample is randomly defined in the range of [0, 1].

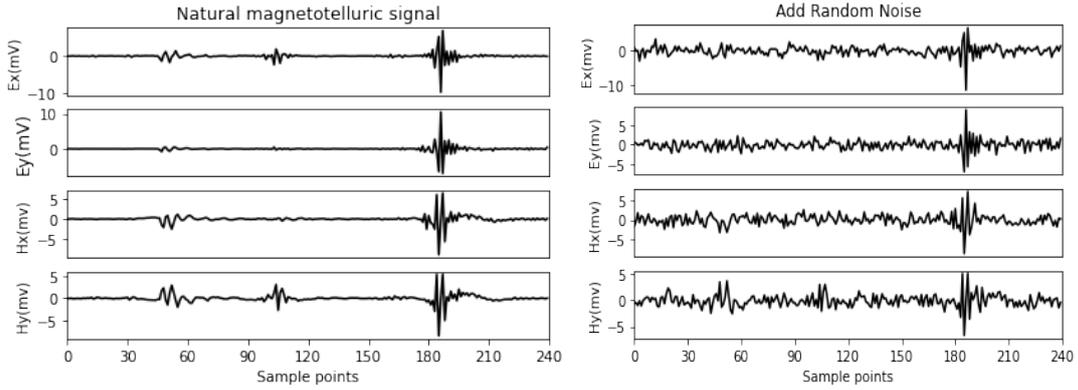

Fig. 4. Comparison of time series: (a) time series without random noise; (b) time series with random noise.

## C. Training Details

As described on the previous section, we generated 30,000 positive samples for training. To avoid any uncertainties associated with the sferics waveform changes between different surveys and acquisition times, each input sample is mapped to a standard normal distribution with mean 0 and variance 1 by the following formula

$$x^* = \frac{x-\mu}{S} \tag{9}$$

where $x^*$ is the normalized input sample, $\mu$ and $S$ are the mean and standard deviation of each channel within the imput sample, respectively. We then preprocess all the input samples by data augmentation discussed before. Without losing the sferics as much as possible, we set the sample length $n = 240$ and the sample mask radius $r = 36$. We train our model with *Adam* optimizer (D. P. Kingma, 2014) and set the parameter $\beta_1 = 0.9$, $\beta_2 = 0.999, \epsilon = 10^{-8}$. The learning rate is initialized to 0.001, and we use a linear learning rate schedule to decrease gradually the learning rate to slow down parameter updates. When the validation metric stagnates for 30 epochs, the initial learning rate is reduced by a factor of 0.5 (Fig. 5c) to achieve a better loss convergence. Moreover, we set batch size as 16, and randomly extract 640 and 160 samples each epoch from training and validation sets. We train our network over 150 epochs, but it can be seen from Fig. 5(a)-(b) that the network saturates early in training, hence we introduce an early stopping strategy to prevent overfitting. Here we set the validation set accuracy as the monitoring indicator of the early-stop strategy, we save the model parameters when the indicator no longer rises within 20 epochs.

Furthermore, we compare the Metrics, Flop counts (GB) and Processing time (s) of different classification models on the validation set and report the results in Table II, which illustrates that the use of data augmentation methods in training can effectively improve metrics, especially the accuracy can be raised by nearly 2%. Among them/Where, VGG19 shows the best results in almost all metrics, and the time cost is relatively cheap, only slightly worse than resnet18. These results are also an important basis for our selection of this network in this research. The training platform in this paper is based on NVIDIA Tesla V100, and it only takes 2.98 seconds to train 1 epoch on VGG19.

We provide the training details in Fig. 5, where Fig. 5(a) presents the loss on training and validation data sets; Fig. 5(b) presents the accuracy on training and validation data sets; and Fig. 5(c) presents the decay curve of the learning rate. We can see that the loss curves for both training and validation gradually converge to less than 0.01 and 0.08, and the accuracy curves gradually stabilize to 97% and 95% after 40 epochs when the optimization stops. According to these data, we can infer that the network already had/got the ability to accurately distinguish between positive and negative samples.

To demonstrate the remarkable performance of the trained network, we first randomly select the time series data of a station in the test set as the input of the network to verify the classification results. The test results are shown in Fig. 6, where Fig. 6(a)-(c) are the model's predictions of positive samples with different S/Ns, and Fig. 6(d) shows a typical negative sample and the model's prediction. It can be seen from the test result that proposed model possessed high accuracy and strong robustness. Even at low S/N, our model is still able to successfully recognize sferic signals from strong background noise level, while simultaneously excluding non-sferic signals. In general, the excellent performance on both the validation and test datasets shows that the network has successfully learned to automatically recognize sferic signals from time series.

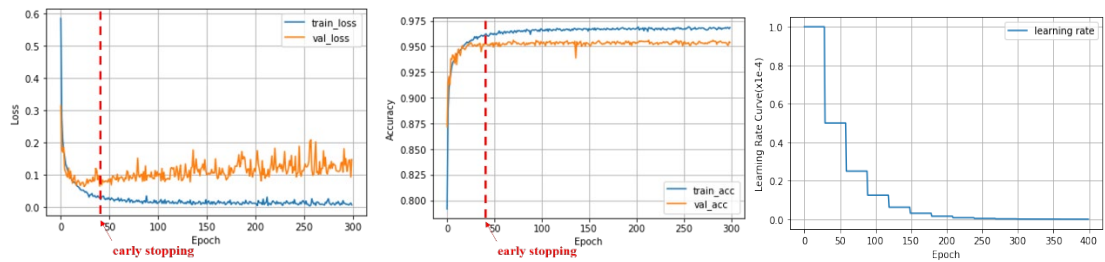

Fig. 5. Training records: (a) Training (blue) and validation (brown) loss curves; (b) Training (blue) and validation (brown) accuracy curves; (c) Learning rate decay curve. The red vertical dashed line indicates the early stopping epochs.

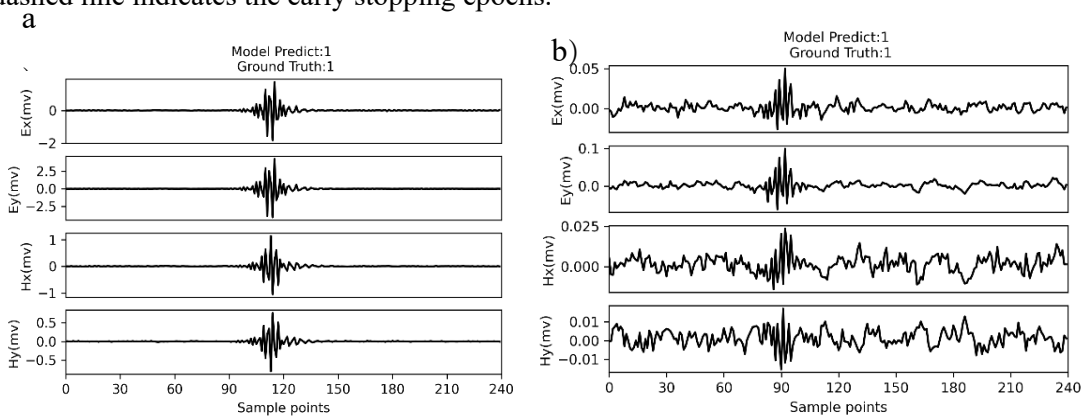

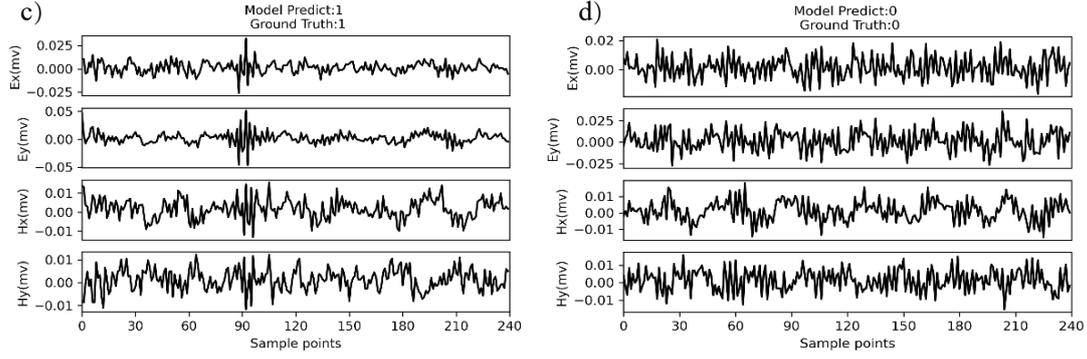

Fig. 6. Model prediction results with different S/Ns: (a) positive samples with high S/N (predict=1, truth=1); (b) positive samples with medium S/N (predict=1, truth=1) ; (c) positive samples with low S/N (predict=1, truth=1); (d) classical negative samples (predict=0, truth=0).

Table II

The Comparison of Different Classification Networks (VGG[11,13,16,19], ResNet[18,34,50,101,152](K He,et al., 2016),VGG without data augmentation, ResNet without data augmentation)

| Network | Metrics | | | | | FLOPs (GB) | Training time(s) /epochs | Predicting time(s)/station |
|---|---|---|---|---|---|---|---|---|
| | BCE | A | P | R | F1 | | | |
| VGG | 0.42(11) 0.30(13) 0.16(16) **0.08(19)** | 0.883(12) 0.914(6) 0.935(4) **0.951(1)** | 0.771(11) 0.824(13) 0.848(16) **0.896(19)** | 0.618(11) 0.718(13) 0.763(16) **0.822(19)** | 0.686(11) 0.767(13) 0.803(16) **0.857(19)** | 1.78(11) 1.92(13) 2.67(16) 3.43(19) | 2.39(11) 2.46(13) 2.57(16) 2.98(19) | 14.3(11) 14.5(13) 14.9(16) 15.6(19) |
| VGG (without data augmentation) | 0.83(11) 0.37(13) 0.27(16) 0.26(19) | 0.863(15) 0.899(10) 0.913(7) 0.936(3) | 0.671(11) 0.786(13) 0.811(16) 0.875(19) | 0.569(11) 0.654(13) 0.673(16) 0.777(19) | 0.615(11) 0.713(13) 0.735(16) 0.823(19) | | | |
| ResNet | 0.28(18) 0.20(34) 0.24(50) 0.27(101) 0.34(152) | 0.918(5) 0.937(2) 0.904(9) 0.872(14) 0.840(17) | 0.843(18) 0.887(34) 0.788(50) 0.726(101) 0.649(152) | 0.739(18) 0.791(34) 0.654(50) 0.583(101) 0.491(152) | 0.787(18) 0.836(34) 0.714(50) 0.646(101) 0.559(152) | **0.32(18)** 1.38(34) 3.11(50) 6.08(101) 8.71(152) | **2.00(18)** 2.19(34) 2.79(50) 5.14(101) 6.77(152) | **14.1(18)** 14.4(34) 15.3(50) 32.4(101) 47.6(152) |
| ResNet (without data augmentation) | 0.46(18) 0.25(34) 0.28(50) 0.30(101) 0.46(152) | 0.896(11) 0.910(8) 0.881(13) 0.853(16) 0.825(18) | 0.784(18) 0.810(34) 0.737(50) 0.661(101) 0.571(152) | 0.626(18) 0.656(34) 0.586(50) 0.520(101) 0.471(152) | 0.696(18) 0.724(34) 0.652(50) 0.582(101) 0.516(152) | | | |

*Bz=16，channels=4，samples_len=240，train_samples=640,val_samples=160.

A=accuracy, P=precision, R=recall, F1=f1score

*The above are the results of the average experiment on the validation set.*

*Numbers in parentheses represent different versions*

*For all metrics except BCE loss, larger values indicate better performance, with the best results highlighted in red.*

# III. RECOGNIZED SFERIC SIGNALS FOR IMPEDANCE ESTIMATION

Before calculating impedance, a preprocessing of time series data is usually required. To avoid sferics from different storm systems with different waveforms contaminating data within a predicted sferic window, we perform waveform correlation filtering for each sferic ensemble, and those with correlations less than a threshold (0.7) will be discarded (Hennessy, 2018). Moreover, time shifting each sferic to align with the ensemble average waveform to correct timing errors. Finally, in order to compare with traditional impedance estimation methods, we use a robust M-estimator to calculate apparent resistivity and phase of the AMT field data preprocessed by our proposed method. The following sections of this part is a brief description of the basic principles and procedures of AMT impedance estimation.

**A. Transfer Functions**

The AMT data analysis starts with estimation of transfer functions between electric field $E$ and magnetic field $H$ (i.e., the impedance tensor) (Sims et al., 1971; Vozoff, 1972).

A common practice is to perform a Fourier transform on the AMT field channel time series $(E_x(t), E_y(t), H_x(t), H_y(t))$ to obtain frequency domain data $(E_x(\omega), E_y(\omega), H_x(\omega), H_y(\omega))$. When the electromagnetic field is a plane wave or zero-wavenumber model, the electromagnetic field component satisfies the following dual-input, dual-output linear system:

$$\begin{pmatrix} E_x \\ E_y \end{pmatrix} = \begin{bmatrix} Z_{xx} & Z_{xy} \\ Z_{yx} & Z_{yy} \end{bmatrix} \begin{pmatrix} H_x \\ H_y \end{pmatrix} \tag{10}$$

When there is noise, it can be written in the following vector form:

$$E = HZ + \varepsilon \tag{11}$$

where $E$ and $H$ denote the horizontal components of the electric and magnetic fields, respectively; $Z$ denotes the impedance tensor, which can be regarded as a transfer function to be estimated in $[mV/(km \times nT)]$; $\varepsilon$ denotes the errors. Afterwards, $\mathcal{O}$ denotes an operator, for example, $Z \simeq \hat{Z} = \mathcal{O}(E, H)$. $r$ denotes the residual, and calculated from:

$$r = E - H\mathcal{O}(E, H) \qquad (12)$$

The definition of the operator $\mathcal{O}$ is not unique in the presence of noisy data. The next section describes the definition and computation of the operator on M-estimator.

**B. M-estimator**

To minimize the effect of data related to large $r$ in regression, Egbert & Booker (1986) introduced a robust technique, called M-estimator, which is considered as one of the most efficient transfer function estimation algorithms (Chave & Thomson, 1989). As long as one remote station is not polluted by correlated noise, it can produce unbiased AMT estimates.

The basic idea of M-estimator is to achieve robust estimation by controlling the influence function, so as to reduce the influence of extreme outliers on the estimation results. Compared with the Standard Least Squares estimation, M-estimator can automatically weight the observed data to reduce the influence of outliers on impedance estimation. Mathematically, the impedance $\hat{Z}$ is defined by following relation:

$$\hat{Z} = \mathcal{O}_M(E, H) = (H^T W(\hat{Z}) H)^{-1} (H^T W(\hat{Z}) E) \qquad (13)$$

where $\mathcal{O}_M$ is the nonlinear weighted least squares operator; $W(\hat{Z})$ is a weighted diagonal matrix that depends on the residuals, given by

$$x(\hat{Z}) = \frac{1}{\beta}(E - H\hat{Z}) \qquad (14)$$

$$W_{j,j}(\hat{Z}) = W(x_j(\hat{Z})) \qquad (15)$$

where $\beta$ is the scale value, which determines the value of the residual that needs to be weighted down; $W$ is a weight function whose purpose is to weaken the influence of large residuals.

In practice, impedance $\hat{Z}$ can be estimated iteratively by

$$\hat{Z}^{i+1} = (H^T W(\hat{Z}^i) H)^{-1} (H^T W(\hat{Z}^i) E), i \geq 0 \qquad (16)$$

The above equation converges when the weighted residuals sum of squares ($r^T r$) changes below a custom tolerance (i.e., 1%). However, the equation does not necessarily guarantee convergence in general, but there are some weighting functions $W$ which can make $\hat{Z}$ converge robustly independently from initial value.

Two key parameters of the M-estimator method: the scale value $\beta$ and the weighting function $W$. Firstly, the scale value $\beta$ must be estimated stably, and Chave et al. (1987) discussed the estimation method of $\beta$

$$\hat{\beta} = \frac{S_{median}}{\sigma_{median}} \qquad (17)$$

where $S$ and $\sigma$ are the sample and theoretical values of the median absolute deviation (MAD), respectively

$$S_{median} = median(|r^{i=0} - media(r^{i=0})|) \qquad (18)$$

When the residual $r$ presents a chi-square distribution, $\sigma_{median} = 0.44845$; and when it presents a normal distribution, $\sigma_{median} = 0.6745$.

There are different weighting functions to choose from in M-estimator, the common ones are Huber weighting function and Thomson weighting function (Holland & Welsch, 1977)

$$W_H(x) = \begin{cases} 1 & |x| < x_0 \\ x_0/|x| & |x| > x_0 \end{cases}, \quad x_0 = 1.5 \qquad (19)$$

$$W_T(x) = e^{-e^{x_0(|x|-x_0)}}, \quad x_0 = 2.8 \qquad (20)$$

Robustness and stability determine the weight function $W$. Compared with the Huber function, the Thomson function is more robust but does not guarantee stability. Therefore, we here use the M-estimate with the Huber weighting function as a good initial value for the Thomson weighting function to overcome the instability.

In summary, the basic principle of the impedance M-estimate is to use iterative weighted least squares to estimate the regression coefficient, then determine the weight through the residual and scale of the previous step, and iterate repeatedly to improve the weight coefficient until the variation of the residuals is lower than a tolerance. Specifically, the procedure for calculating the impedance M-estimate with the Thomson weighting function is as follows

(1) Calculate least-square estimator $\hat{Z}_{OLS}$

(2) Calculate residual $r_{OLS}$ and scale $\beta_{OLS}$

(3) Calculate M-estimator with Huber weight

(4) Calculate residual $r_H$ and scale $\beta_H$

(5) Repeat step (3) if the variation of residuals $\Delta r > tol(1\%)$

(6) Calculate M-estimator with Huber/Thomson weight

(7) Calculate residual $r_H/r_T$ and scale $\beta_H/\beta_T$

(8) Repeat step (6) if the variation of residuals $\Delta r > tol(1\%)$

(9) Compute impedance $\hat{Z} = (H^T W(\hat{Z})H)^{-1}(H^T W(\hat{Z})E)$

## C. Discrete Fourier Transform Computation

The field channel time-series data of duration $T(s)$ can be expressed as the following relationship

$$T = \gamma N_w \frac{N_p}{F} \tag{21}$$

where, $N_w$ denotes the equally sized and evenly spaced time windows split from the time series; $F$ denotes the target frequency, $N_p$ denotes the number period, then the duration of one window can be expressed as $\frac{N_p}{F}$; $\gamma$ denotes the window overlap ratio, and this shifting can increase $N_w$ or conversely decrease the correlation between windows.

Subsequently, the number of windows can be determined by

$$N_w = \frac{TF}{\gamma N_p} \tag{22}$$

When time series are divided into $N_w$ time portions, we can use the Slepian data taper windows with time bandwidth $\tau$ ($\tau=1,2,3$ or $4$) to calculate DFTs.

## IV. FIELD DATA APPLICATIONS

Our CNN model achieved robust results on most field data sets, even if it is only trained with data from fewer surveys. In this part, to verify the capability of the model, we feed two field data from different surveys into the well-trained CNN model. After considering the geological conditions, field sources, noise sources and other factors, we finally choose Nanjing survey with strong background culture noise and Wuhan survey with complex EM and anthropogenic noises to make up the test set. To be consistent with training, each input sample is normalized as same as the training sample, but without data augmentation.

In order to analyze the processing effect of proposed method conveniently and intuitively. For stations, we mainly compare the improvement of power spectrum, apparent resistivity and phase curves in AMT dead-band. The property of AMT compels that the observed response function on the Earth's surface must vary smoothly with frequency, and we cite this increase in smoothness as strong evidence of the superiority of the data processing method (Garcia, 2005; Booker, 2014). The phase tensor ellipse graphically reflects how the phase relationship varies with polarization, with the major axis of the tensor represented by the major and minor axes of the ellipse (Caldwell, 2004). For survey, the evaluation indicators selected are mainly the phase tensor pseudo-section and the apparent resistivity and phase pseudo-section in the dead-band. Phase tensor pseudo-sections provide information about the directionality of the regional resistivity structure. Apparent resistivity and phase pseudo-section can more intuitively reflect the subsurface resistivity distribution. The above-mentioned indicators are the main parameters of distortion (Hennessy, 2017).

**A. Case Study One—— Nanjing**

The first field dataset was acquired from Nanjing city, Jiangsu province, China. Gaochun District is located at the southwest end of Jiangsu Province, adjacent to Liyang and Lishui in the east and north respectively, and borders Dangtu, Xuanzhou and Langxi in Anhui Province in the southwest. On the tectonic unit, it belongs to the edge of the Nanjing Sag. The anticline of the Ligao fold-uplift belt slants through the area from north to south. The eastern hills are ups and downs. The survey is far away from the city and has less strong human interference which mainly exists in the form of underground stray current.

Fig. 7 displays a time series segment of station01 of Nanjing survey. As shown, we can see that the proposed method can effectively recognize sferics from relatively clean time series, which proves that our model indeed learns the feature representation about the waveform of sferics. It can be seen from the data in Table III that each metric also achieves great results on the test set that our model has never seen before, which strongly demonstrates the excellent generalization of our model.

Fig. 8(a)-(b) show the power spectrum comparison of station 001. As can be seen from Fig. 8(a) that there are Industrial current interferences (fundamental frequency is 50hz) in the original time series, which appear as fixed frequency peaks on the power spectrum. Fig. 8(b) verifies that our approach is able to suppress strong noise to restore real data spectrum, and to better highlight the main frequencies (1khz, 1.8khz, 2khz, 2.5khz, 3.9khz and 4.5khz) of sferic signals to compensate for the lack of natural field energy in the dead-band (1.5 khz-5khz)

Fig. 9(a)-(b) compare the apparent resistivity and phase curves of three adjacent stations 01-03 between 700 Hz and 10 kHz, where the point distance is 20m. For the stations, Fig. 9(a) shows that the apparent resistivity $\rho_{xy}$ and $\rho_{yx}$ decays abnormally with a degree of two orders of magnitude between multiple frequency points at dead-band frequencies, which obviously violates the objective law that the underground resistivity structure must vary slowly. And the abnormal deviation of the phase curve and the confidence interval of the error bar indicate a serious impedance estimation error in AMT dead-band.

Fig. 9(b) shows the apparent resistivity and phase curves calculated using our method. The apparent resistivity $\rho_{xy}$ and $\rho_{yx}$ recovers to its normal value, and three curves are smoother and have smaller standard errors, restoring better consistency between adjacent stations. Correspondingly, the phase tensor ellipse also shows great depolarization properties, with its major axis recovering the lateral variation of the real subsurface resistivity structure. All the findings above verify the effectiveness of our method.

For this survey, Fig. 10(a) shows that there is severe electric field distortion on the original phase tensor pseudo-section, which usually changes the polarization direction of the local electric field, thus causing the observed response is distorted by the local conductivity inhomogeneity (Caldwell, 2004). By computing the phase tensor for all 26 combinations of synchronous stations, as shown in Fig. 10(b), the pseudo-section preprocessed by our method exhibits smoother frequency variations at dead-band frequencies, while largely eliminating the distortion of the observed impedance tensor, and recovering the graphical representation of the tensors involved in the galvanic distortion of a 2D impedance tensor.

Fig. 11(a) shows that a common problem in this survey, that is, the lack of AMT field energy leads to a misestimation of impedance in the dead-band. Specifically, both apparent resistivity $\rho_{xy}$ and $\rho_{xy}$ present a low resistance between $10^{-4}$(s) and $10^{-3}$(s), which is shown as a dark brown long axis along the transect, and the occurrence of rare anomalous phase behavior (phase reversal) at 56, 61, 66 stations. Comparing Fig. 11(a) highlights the superiority of our proposed method between 700 Hz and 10 kHz. The apparent resistivity-phase pseudo-sections in Fig. 11(b) tends to be smooth, and the anomalous behavior of low resistance in the dead-band and the estimation error of the phase reversal are corrected, which further restores the real subsurface resistivity structure.

Although the structural patterns and waveform features are different from the training data. But overall, the results above indicate that our method works well on this field data example and show a significant improvement in AMT dead-band.

Table III
Quality metrics of test stations in Nanjing survey

| Station | 001 | 002 | 003 | 004 | 005 | 006 | Mean |
|---|---|---|---|---|---|---|---|
| A | 0.969 | 0.956 | 0.959 | 0.932 | 0.946 | 0.936 | 0.949 |
| P | 0.723 | 0.849 | 0.717 | 0.791 | 0.727 | 0.841 | 0.774 |
| R | 0.96 | 0.756 | 0.962 | 0.782 | 0.652 | 0.667 | 0.796 |
| F1 | 0.825 | 0.800 | 0.822 | 0.786 | 0.687 | 0.744 | 0.777 |

*A=accuracy, P=precision, R=recall, F1=f1score

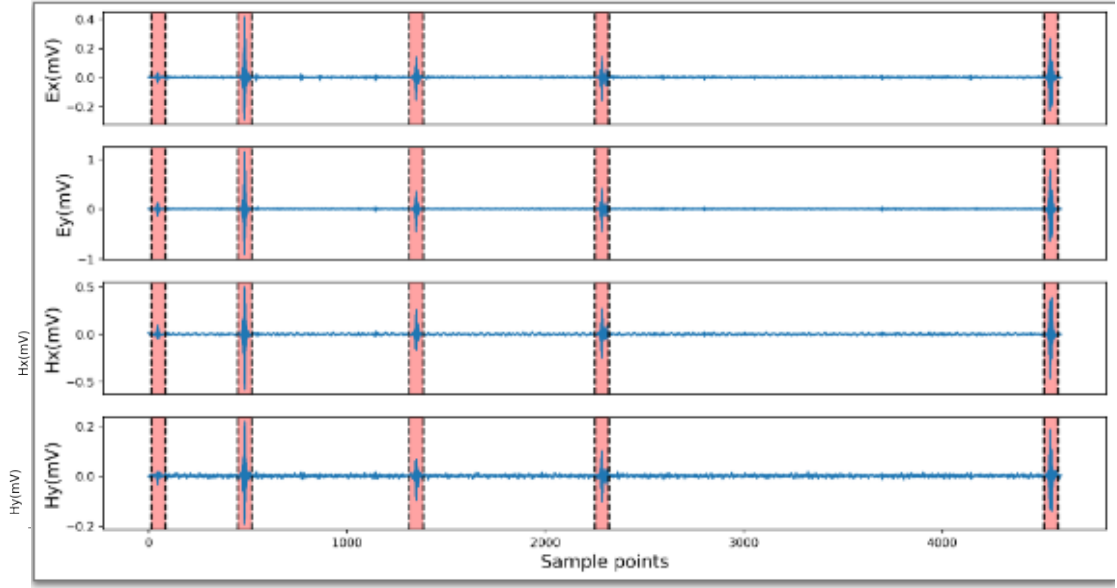

Fig. 7. A time-series segment of Nanjing. The pink part indicates the sferics recognized by our network

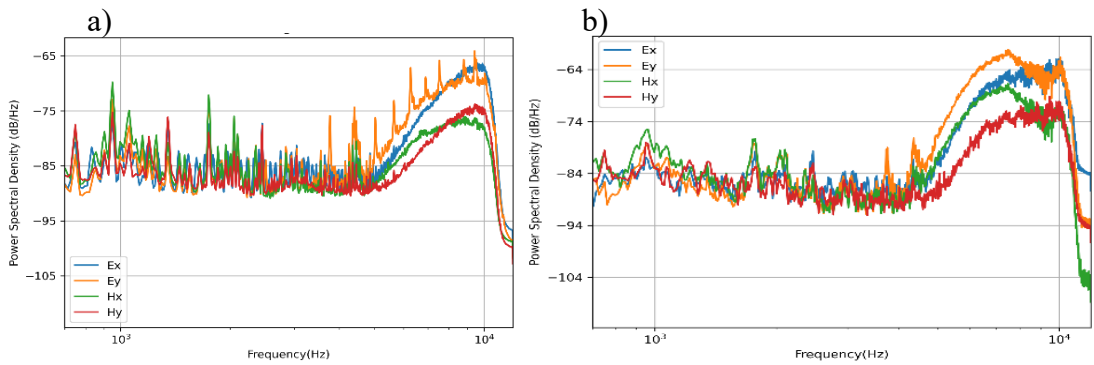

Fig. 8. Comparison of power spectrum (700Hz-10400Hz) for station01: (a) results calculated by conventional methods; (b) results calculated by our workflow

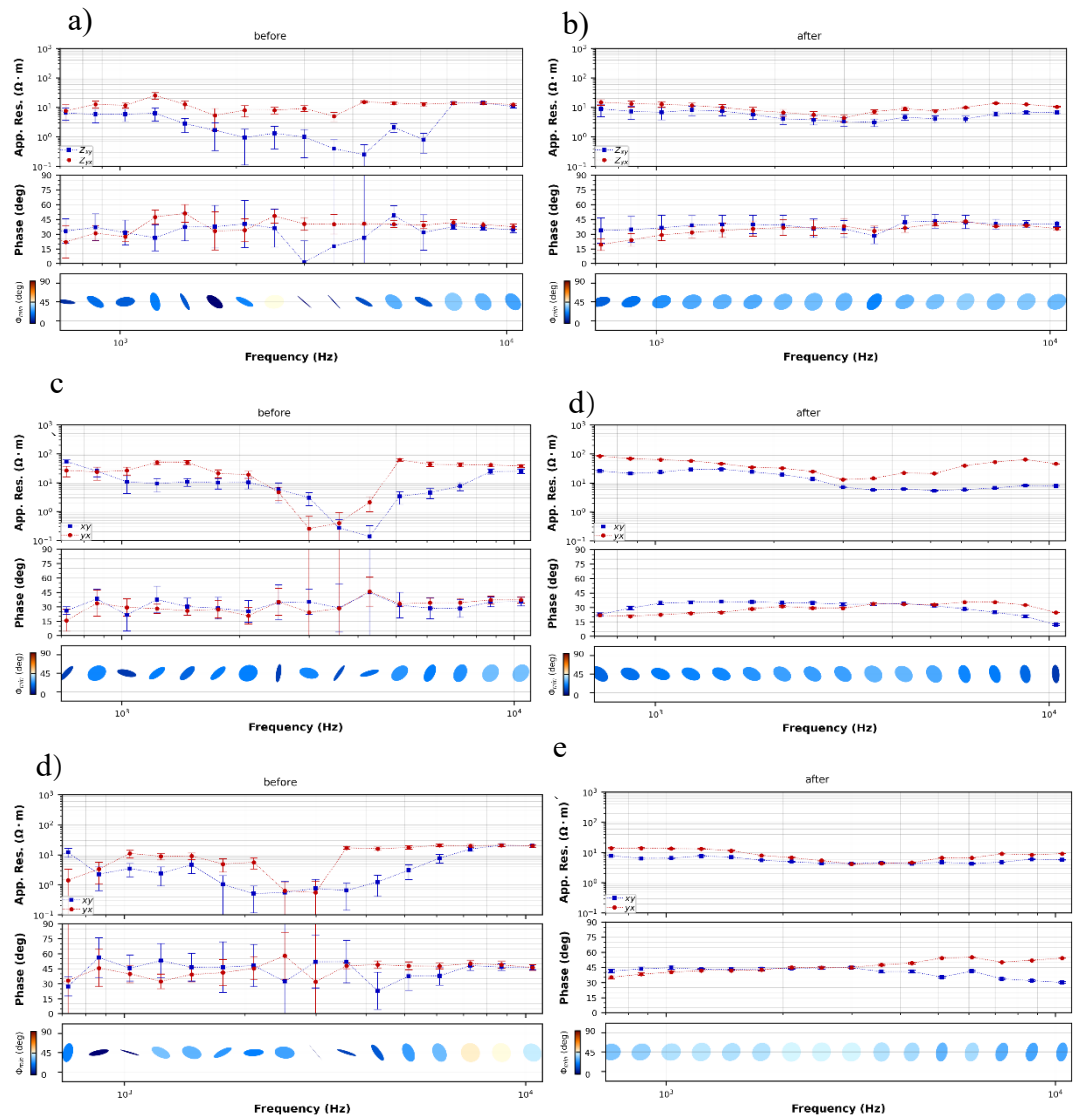

Fig. 9. Comparison of apparent resistivity-phase curves and phase tensor ellipse (700Hz-10400Hz) for station01(top), station02(middle) and station03(bottom): (a) results calculated by conventional methods; (b) results calculated by our workflow

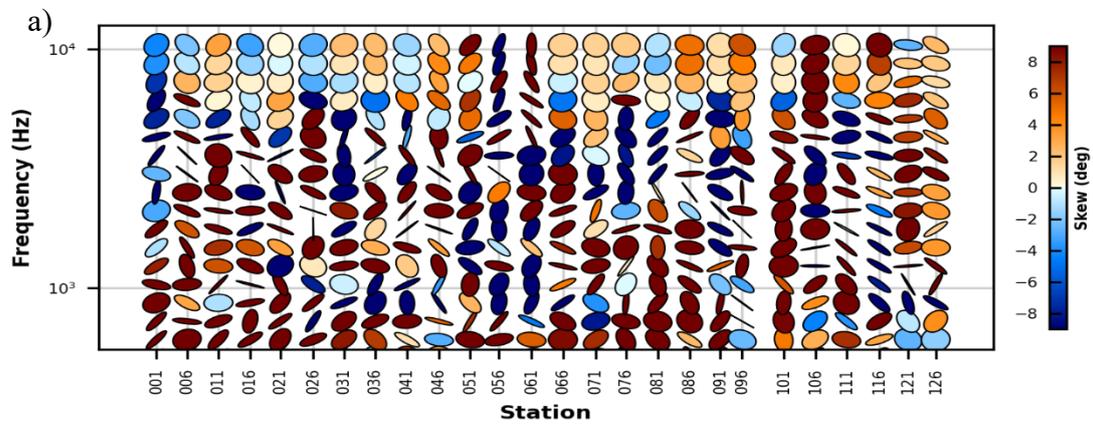

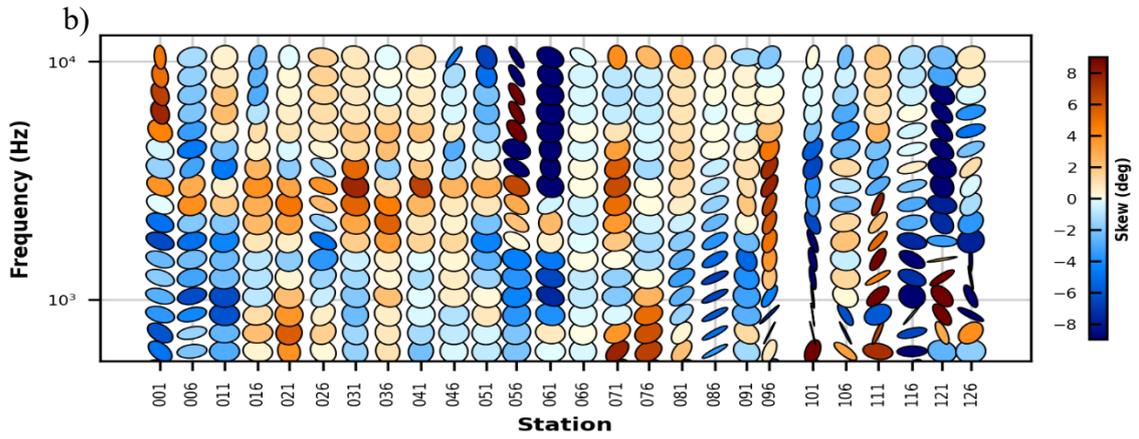

Fig.10. Comparison of phase tensor pseudo-sections (700Hz-10400Hz) in Nanjing, the color bar is limited to [-10° 10°] range. (a) results calculated by conventional methods; (b) results calculated by our workflow

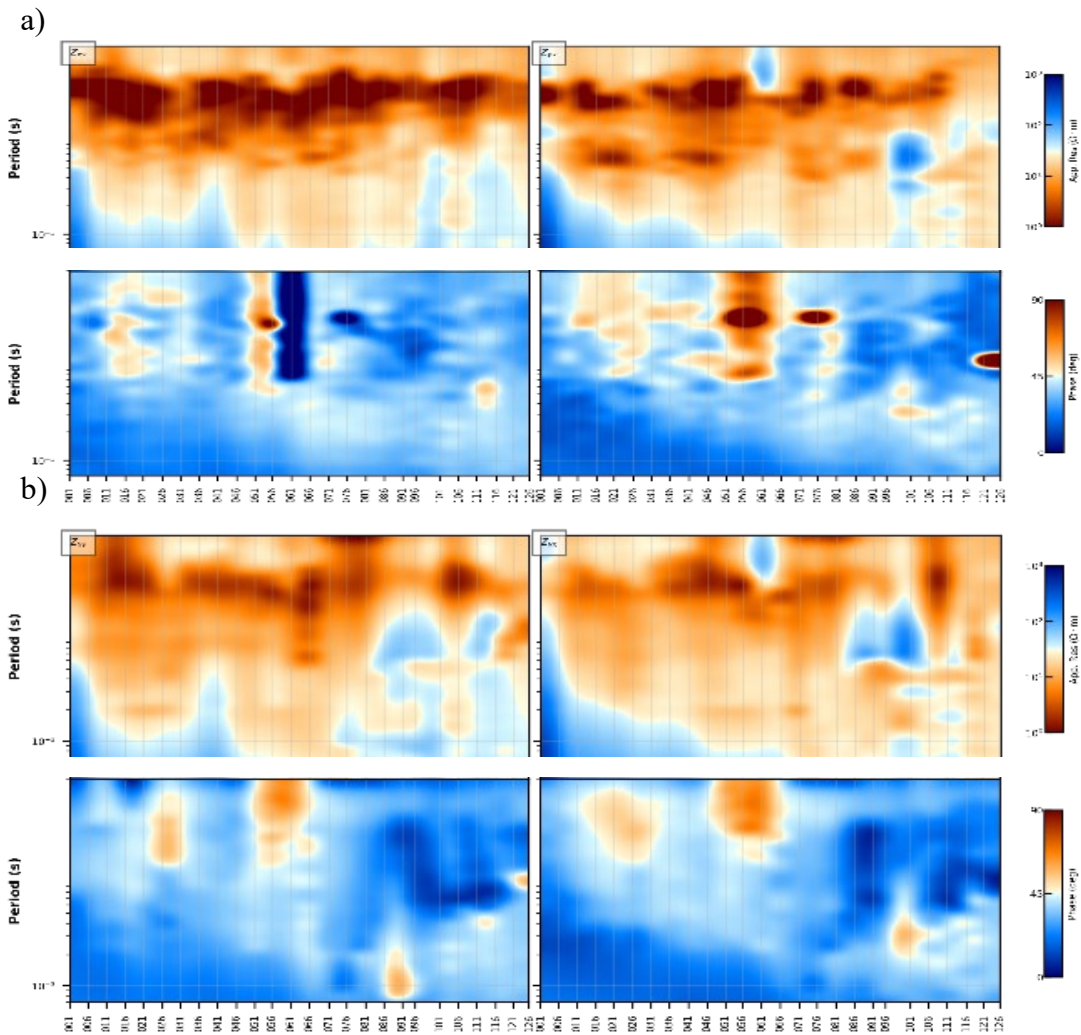

Fig. 11. Comparison of apparent resistivity-phase pseudo-sections (700Hz-10400Hz) in Nanjing, we use the bicubic interpolation method to give a smoother interpretation. (a) results calculated by conventional methods; (b) results calculated by our workflow

## B. Case Study Two—— Wuhan

To further verify the reliable generalization of our model, we add the second real data example. The second field data comes from Wuhan city, Hubei province, China. The landform of this area is a transitional area from the east margin of Hanjiang Plain to the south foot of Dabie Mountain, with low hills in the middle, surrounded by hills and ridges in the north and south. Located at the junction of mountains and cities, the survey is rich in industrial noise (electrical wires, factories, electrical equipment and so forth) generated by electromagnetic fields associated with human activities and geological noise caused by changes in topography and the presence of shallow inhomogeneous bodies (relief, randomly distributed rock masses). During the exploration, there multiple artificial source geophysical methods (i.e., CSEM, CSAMT, DFIP, etc.) working at the same time, which make the time-series waveform more complex. In fact, the second real data example is the most challenging one, and robustly obtaining accurate apparent resistivity and phase of this field data remains a challenge for existing methods.

Fig.12 shows that field data acquired at station41 present more energetic background noise interference in the time series, the existence of these disturbances brings great challenges to data interpretation in the dead-band. What is surprising is that our well-trained model is able to accurately recognize sferic signals even in the face of such complex time-series waveforms. This finding further supports the idea of the strong expressiveness of our CNN model and verifies the robustness of model against data noise. Table IV shows the classification quality metrics on the test set of Wuhan survey. Affected by low S/N data, the average accuracy of model classification in Wuhan survey is nearly 2% lower than that of Nanjing survey.

Fig. 13(a)-(b) show the power spectrum comparison of station41. Fig. 13 (a) shows that the station is severely disturbed by cultural interference, with the noise signal predominant from 1khz to 6khz, particularly. Fig.13(b) shows that after processing by our method, the real frequency of the nature field can be basically recovered from the power-line/industrial interference. However, due to the few sferic events acquired and the existence of strong interference sources (i.e., power lines), we can still see that there is a lack of field source energy and a few residual harmonic disturbances in the dead-band.

Fig.14(a)-(b) shows the apparent resistivity and phase comparison results of station41, station46 and station50, and comparison results demonstrate that low S/N in the dead band cause poorly estimated apparent resistivity and phase. Fig.14(a) shows the lack of smoothness for both $\rho_{xy}$ and $\rho_{yx}$ at the three stations. The difference between the minimum and maximum of the curve is close to 4 orders of magnitude, error bars and phase tensor ellipses also indicate a breakdown of robust impedance estimation methods caused by strong noise interference, especially in station46. Fig.14(b) shows the results after processing by our method. It can be seen that the above problems have been properly resolved. The apparent resistivity curves eliminate spurious electrical structural changes and restores smoothness, and the phase curve corrects the abnormal offset to restores good consistency, although the error bar and phase tensor ellipse still slightly affected by interference.

Fig. 15(a) shows that the original data is severely disturbed by polarization along the transect, especially in station26-34, station41, station43, and station46. Fig.15(b) shows that the phase tensor pseudo-section is partially recovered from the distortion of the phase tensor ellipse caused by strong noise interference, but there is still room for further improvement. From Fig. 16(a), we can see the obvious anomalies of apparent resistivity mutation and phase reversal, which is specifically manifested in the above-mentioned stations. Fig. 16(b) shows that our method is still effective in the face of low S/N data, and the processed pseudo-section is smoother and more reasonable, which is consistent with real subsurface resistivity structure. Finally, the anomalies present in Fig.16(a) are well corrected.

For both two field data examples, we never construct specific training samples with similar noise source types and waveform characteristics. However, model trained only with original time series data still shows remarkable performance by accurately recognizing the large amplitude sferic signals contained in the time series, although the extracted sferic signals may not strictly match the hand-picked ground truth. These performances imply that the trained model has successfully learned global feature patterns, rather than simply remembering the input training samples.

Table IV
Quality metrics of test stations in Wuhan survey

| Station | 01 | 02 | 03 | 04 | 05 | 06 | Mean |
|---|---|---|---|---|---|---|---|
| A | 0.946 | 0.931 | 0.949 | 0.92 | 0.897 | 0.916 | 0.926 |
| P | 0.643 | 0.512 | 0.678 | 0.543 | 0.501 | 0.507 | 0.564 |
| R | 0.873 | 0.886 | 0.865 | 0.923 | 0.916 | 0.927 | 0.898 |
| F1 | 0.74 | 0.649 | 0.76 | 0.684 | 0.648 | 0.655 | 0.689 |

*A=accuracy, P=precision, R=recall, F1=f1score

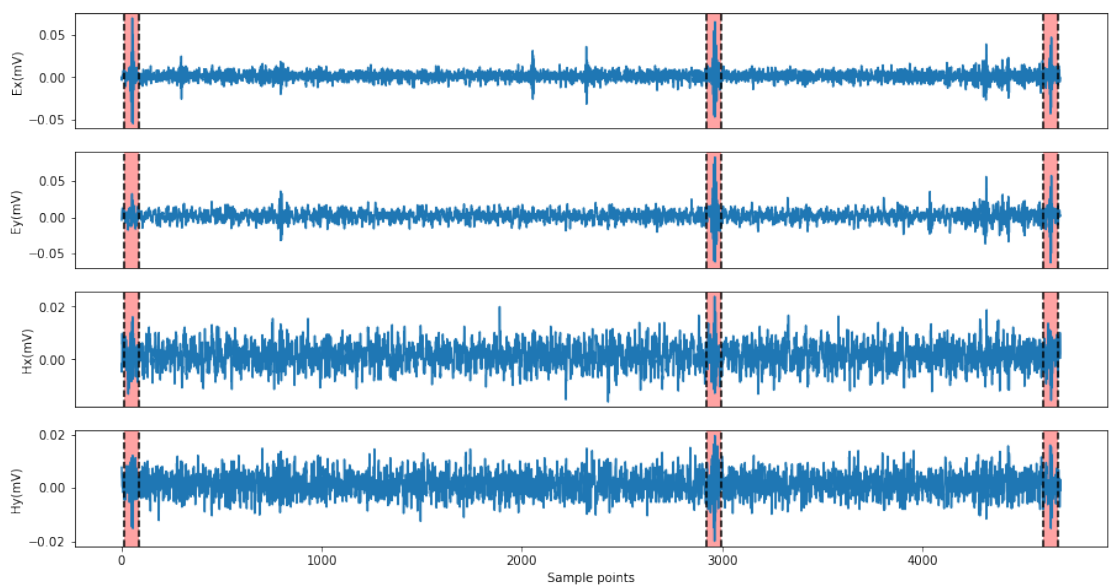

Fig. 12. A time-series segment of Wuhan. The pink part indicates the sferics recognized by our network

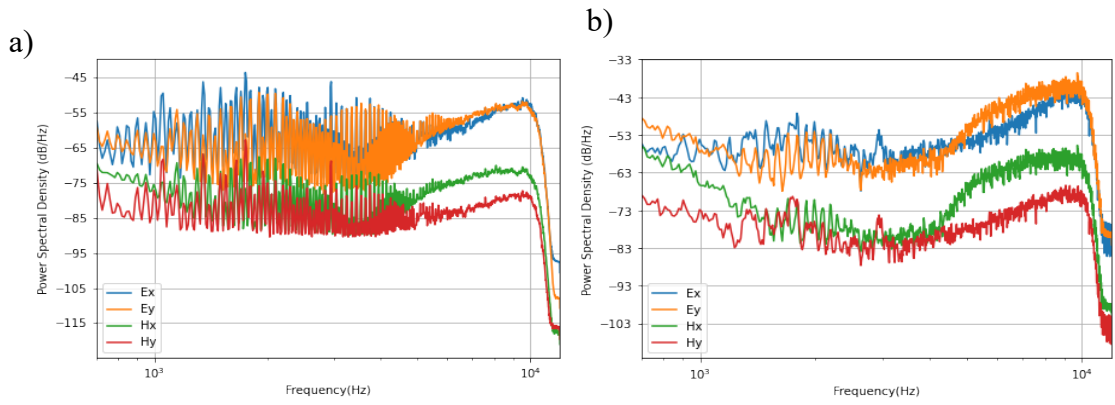

Fig. 13. Comparison of power spectrum (700Hz-10400Hz) for station41: (a) results calculated by conventional methods; (b) results calculated by our workflow

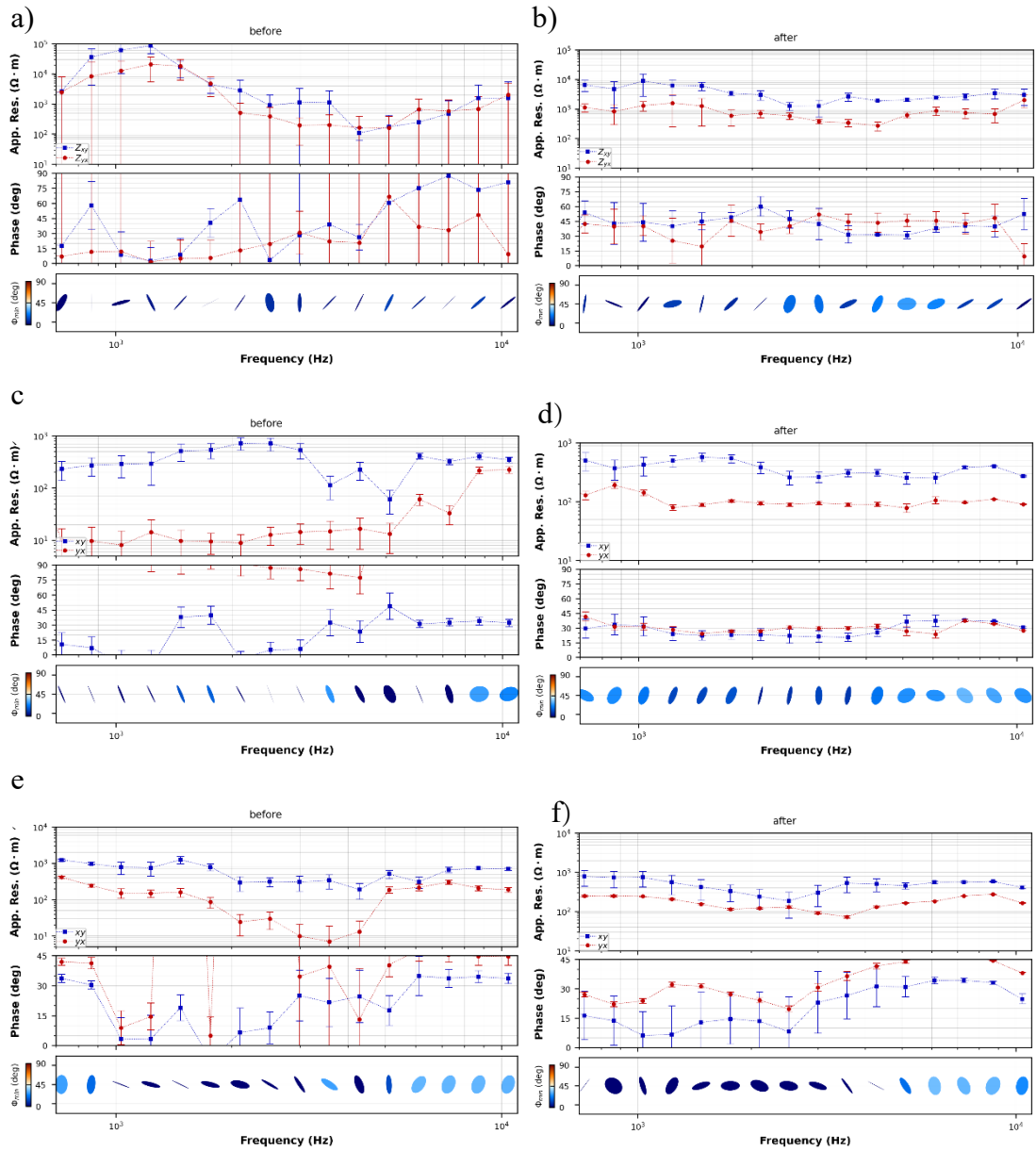

Fig. 14. Comparison of apparent resistivity-phase curve and phase tensor ellipse (700Hz-10400Hz) for station41(top), station46(middle) and station50(bottom): (a) results calculated

by conventional methods; (b) results calculated by our workflow

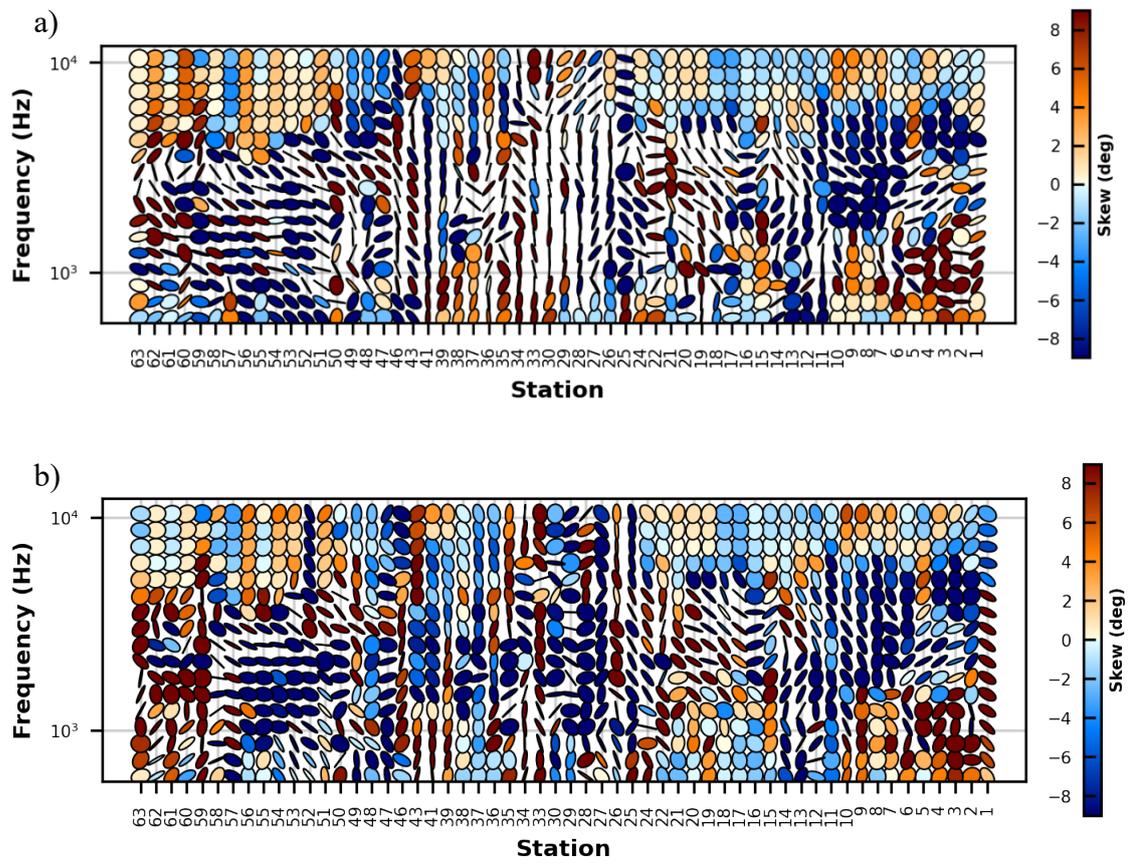

Fig. 15. Comparison of phase tensor pseudo-sections (700Hz-10400Hz) in Wuhan. (a) results calculated by conventional methods; (b) results calculated by our workflow

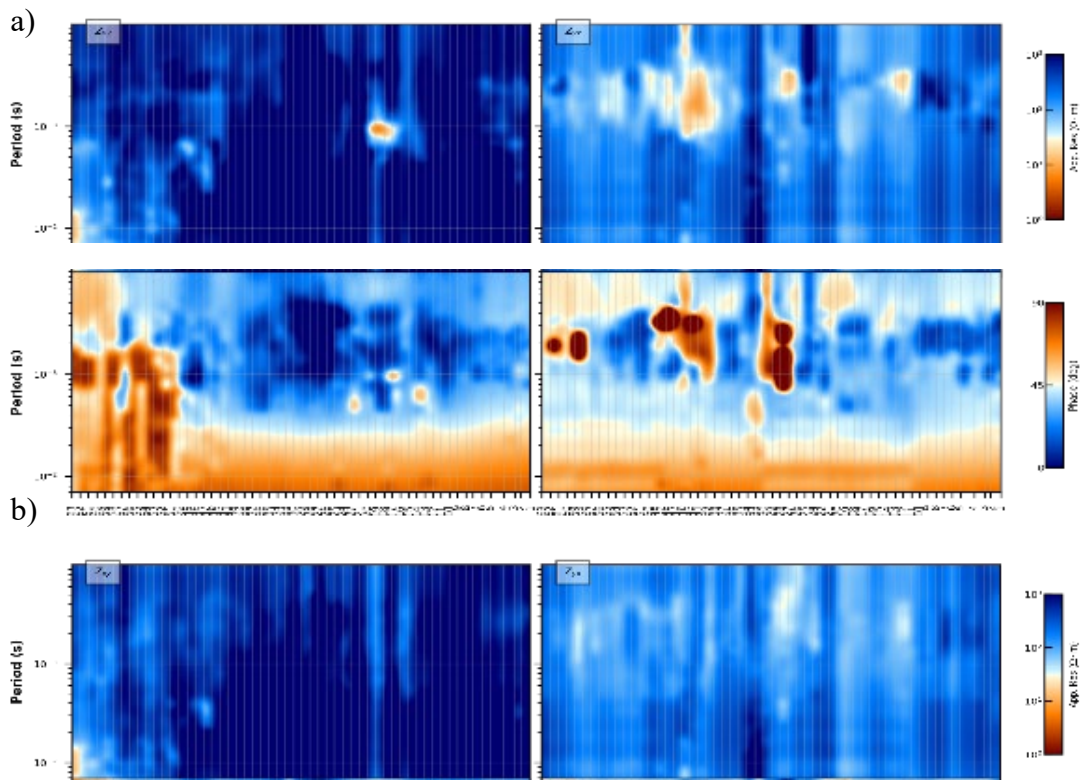

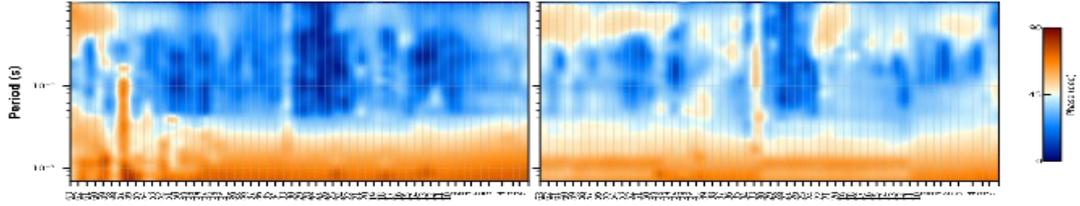

Fig. 16. Comparison of apparent resistivity-phase pseudo-sections (700Hz-10400Hz) in Wuhan. (a) results calculated by conventional methods; (b) results calculated by our workflow

## V. DISCUSSION

During forward prediction, we found that the model failed to recognize the positive samples whose sferics center point $ps$ was at the edge of the sampling window. A possible explanation for this might be that when making positive samples, in order to make the sampling window contain complete sferics, the center point $ps$ of each sferic is set to be at least $r$ sampling points away from the edge of the window. Therefore, the above-mentioned positive samples are lacking in the given training set, which eventually leads to misjudgment in the forward prediction. To solve this problem, we use a sliding window with an overlap ratio of 50% in the forward prediction, which enables sferics center point $ps$ to shift from the edge of the sampling window to the center of the sampling window. As it turns out, this strategy achieves a much lower probability of false positives at an acceptable time cost.

The method proposed in this paper showed good generalizability on datasets with different S/Ns, but it is essentially data-driven and limited by the number and quality of training samples. When the number and amplitude of sferics contained in the time series are extremely low, and noise fills the entire time series to drown out the sferic signals, the method breaks down. The generated samples contaminated with strong noise are not enough to calculate robust apparent resistivity and phase results, resulting in severe distortions in the interpretation of AMT data. Therefore, if there is a large interference in the acquisition time or within the survey area, we recommend using a variety of advanced exploration methods and anti-interference algorithms for preprocessing to obtain a high S/N time series. Theoretically, the results achieved using hybrid methods will be better than using our method only.

In view of the fact that our model has achieved a high sferics recognition accuracy on the existing dataset, in the follow-up work, we plan to use semi-supervised learning methods to automatically label data, and acquire generally representative data to enlarge the training set. In this way, we can further reduce the time cost of manual annotation and achieve high-precision recognition of signals under complex conditions.

The training parameters used in this paper are not optimal, and we strongly recommend setting your own parameters tuning strategy according to the actual situation.

## VI. CONCLUSION

The main goal of the current study is to obtain superior impedance estimation in the AMT dead-band. In this paper, we focus on the recognition and extraction of sferics, and provides a simple and effective method. Specifically, we propose a novel CNN-based method for sferic signals recognition. Without any manually interception of time series, the well-trained network can automatically recognize and extract sferic segments from original time series. The proposed network is a one-dimensional variant of VGG, and we employ random sampling windows to generate samples, add random noise for data augmentation, and finally use weighted BCE as the loss function to optimize network parameters during training.

The application of two field data examples verifies that the trained CNN model not only performs well on data not included in training, but also reached excellent robustness and generalization on data with different S/Ns. Multiple examples demonstrate that proposed method effectively solves the problem of lack of energy in the AMT dead-band, also eliminates the current distortion effect displayed by the phase tensor pseudo-section, and corrects the abnormal distortion present in the apparent resistivity-phase pseudo-section. By using our proposed method, we obtained smoother apparent resistivity-phase curves at dead-band frequencies (1.5–5 kHz), which restores the real subsurface resistivity structure. These results of this study have a number of important implications for future practice such as from mineral resource exploration to geothermal energy production.

However, the generalizability of these results is subject to certain limitations. The main limitation of the proposed method discussed in this study is the S/N of the time series data. When data S/N is low, the sferic segments containing strong noise extracted by our method will reduces the stacking window of the impedance estimation but without improving the data S/N. Further research should be undertaken to explore how to properly combine advanced recognition and denoising algorithms to achieve high-precision extraction of field source main signals. Furthermore, we want to complicate our data generation workflow by increasing the diversity of sferic waveforms to obtain more realistic waveform characteristics and data distributions. By doing so, it is expected to improve generalization on more field data examples and refine the overall workflow.

## VII. ACKNOWLEDGEMENTS

This research was funded by the National Natural Science Foundation of China (42130810), the Scientific research project of Hunan Province Institute of Geology (HNGSTP202201), and the SinoProbe Laboratory (Chinese Academy of Geological Sciences). We acknowledge the use of the Razorback library and MTPy library for magnetotelluric data analysis and visualization.

### Data Availability Statement

The field data used in this study could be obtained by direct request to the corresponding author Rujun. Chen.